\documentclass[10pt,conference]{IEEEtran}
\IEEEoverridecommandlockouts

\usepackage{graphicx}
\usepackage{grffile}
\usepackage{verbatim}
\usepackage[cmex10]{amsmath}
\usepackage{amsthm}   
\usepackage{array}   
\usepackage{mdwmath}   
\usepackage{mdwtab}   
\usepackage{eqparbox}   
\usepackage{times}   
\usepackage{fixltx2e}   
\usepackage{algorithmic}
\usepackage{algorithm}
\usepackage{float}   
\usepackage{color,soul}   
\usepackage{multirow}   
\usepackage{amssymb}   
\usepackage{relsize}
\usepackage{amsfonts}
\usepackage{multicol}
\usepackage{comment}
\usepackage{mathtools}
\usepackage{lipsum}
\usepackage{cuted}
\usepackage{caption}
\usepackage{subcaption}
\usepackage[numbers,sort&compress]{natbib}

\graphicspath{ {./figs/} }

\title{Simulation-Based Analysis of COVID-19 Spread Through Classroom Transmission on a University Campus}%

\author{\IEEEauthorblockN{Arvin Hekmati\IEEEauthorrefmark{1}, Mitul Luhar\IEEEauthorrefmark{3}, Bhaskar Krishnamachari\IEEEauthorrefmark{1}\IEEEauthorrefmark{2}, and Maja Matari\'c\IEEEauthorrefmark{1}\\}
\IEEEauthorblockA{\IEEEauthorrefmark{1}Department of Computer Science \\ 
\IEEEauthorrefmark{2}Department of Electrical and Computer Engineering\\ 
\IEEEauthorrefmark{3}Department of Aerospace and Mechanical Engineering \\ 
University of Southern California\\ Los Angeles, California, USA\\ Email:
  \texttt{\{hekmati,luhar,bkrishna,mataric\}@usc.edu}\\[1ex]}}
 
\usepackage{datetime}
\newdate{date}{01}{31}{2020}
\date{\normalsize{\displaydate{date}}}

\IEEEpubid{\begin{minipage}{\textwidth}\ \\[12pt]
 This work has been supported by a Provost’s New Strategic Directions in \\ Research and Scholarship Award at USC.
\end{minipage}}

\begin{document}

\allowdisplaybreaks[1]
\maketitle

\begin{abstract}

    Airborne transmission is now believed to be the primary way that COVID-19 spreads. We study the airborne transmission risk associated with holding in-person classes on university campuses. We utilize a model for airborne transmission risk in an enclosed room that considers the air change rate for the room, mask efficiency, initial infection probability of the occupants, and also the activity level of the occupants. We introduce, and use for our evaluations, a metric $R_0^{eff}$ that represents the ratio of new infections that occur over a week due to classroom interactions to the number of infected individuals at the beginning of the week. This can be seen as a surrogate for the well-known $R_0$ reproductive number metric, but limited in scope to classroom interactions and calculated on a weekly basis. The simulations take into account the possibility of repeated in-classroom interactions between students throughout the week. We presented model predictions were generated using Fall 2019 and Fall 2020 course registration data at a large US university, allowing us to evaluate the difference in transmission risk between in-person and hybrid programs.  We quantify the impact of parameters such as reduced occupancy levels and mask efficacy. Our simulations indicate that universal mask usage results in an approximately $3.6\times$ reduction in new infections through classroom interactions. Moving 90\% of the classes online leads to about $18\times$ reduction in new cases. Reducing class occupancy to 20\%,  by having hybrid classes, results in an approximately $2.15-2.3\times$ further reduction in new infections. 
\end{abstract}



\section{Introduction and Related Work}

    The COVID-19 pandemic has had a profound impact on educational institutions around the world.  More than 85 colleges and universities across the US have reported at least 1,000 cases of COVID-19, and over 680 institutions have reported at least 100 cases~\cite{nytimes2020track}.  More than 124,000 public and private schools, colleges, and universities in the US closed in April 2020, impacting more than 55 million students~\cite{education2020closure}. Worldwide, similar disruptions have affected more than 1.7 billion students~\cite{duong2020ivory, wiki2020impact}.

    In response to the initial COVID-19 outbreak in Spring 2020, a large number of colleges and universities across the US decided to cancel classes and close student housing~\cite{wiki2020impact}. Many universities and colleges moved instruction online. 
    In Fall 2020, many institutions of higher education in the US returned to in person instruction.  However, this led to a significant increase in new infections. Several colleges and universities decided to reopen in the Fall 2020 and provide hybrid classes where a portion of the students could attend the classes in person while the others attended online, providing a partial solution to the problems associated with purely online instruction. In addition, colleges and universities put in place rules about physical distancing and face covering usage and limited social gatherings.  Many institutions also put in place extensive population testing, contact tracing, and quarantining measures for on-campus students, staff, and faculty. This combination of measures had some success in curbing the spread of COVID-19 on campuses~\cite{nytimes2020track}. 

    
    In this work, we study the impact of different policies on transmission through classroom interactions at universities. We characterize the effect of parameters such as reduced occupancy and mask wearing on the number of new infections generated via university classroom interactions.  In particular, we compare model predictions generated using course registration data from a large US university for Fall 2019, when all classes were held in person, and for Fall 2020, when most classes were online and some were conducted in hybrid mode. 
    
    The main contribution of this work is to \emph{quantify} the impact of university policies on COVID-19 transmission through classroom interactions. Our analysis indicates that, with the Fall 2019 schedule in place, universal mask wearing would have resulted in a roughly $3.6\times$ reduction in new cases and a reduction in classroom occupancy to 20\% would have resulted in a further $2.15-2.3\times$ reduction. Moving 90\% of classes online, as was done for Fall 2020, leads to a roughly $18\times$ reduction in cases relative to the Fall 2019 baseline, with universal masking and reduced occupancy leading to further reductions.  Together, these findings suggest that the precautions taken by US institutions of higher education may have had a significant impact in curbing the spread of COVID-19 via classroom interactions.

    The rest of the paper is organized as follows. Section \ref{sec:general_risk_model} presents a general model for transmission risk in enclosed spaces. In section \ref{sec:classroom_risk_model}, we describe how this model is used to compute transmission via classroom interactions in our simulations. In section \ref{sec:dataset}, we present the Fall 2019 and Fall 2020 datasets that we used in the simulations. The simulation methodology is discussed in section \ref{sec:simulation_methodology}.  Simulation results are presented and discussed in section \ref{sec:results_discussion}. Finally, in section \ref{sec:conclusion} we conclude the paper.
\vspace{-1mm}
\section{General Risk Model}
\label{sec:general_risk_model}
    In this section, we present a simple model for airborne virus emission and exposure in an enclosed space. The goal is to provide an \textit{estimate} for airborne virus concentrations and dosage for a known number of occupants and duration of proximity. Airborne virus concentrations depend on the number of infectious persons in the room, whether the occupants are being active or passive, as well as any mitigating factors such as the use of face coverings, enhanced HVAC protocols, and limited occupant density due to physical distancing. The dosage for exposed individuals further depends on the duration of proximity and the effectiveness of any face coverings. The main assumption of this model is that it considers perfectly-mixed conditions in the room, which means the concentration of virus particles is uniform.  In simple terms, the model assumes that any airborne particles are mixed throughout the space quickly. This assumption implies a uniform transmission risk for all occupants in the room.  Such \textit{mixed flow} or \textit{continuously-stirred reactor} models are common in indoor air quality modeling \cite{nazaroff2004indoor,nazaroff2016indoor}.
    In the context of airborne disease transmission, such models are typically referred to as Wells-Riley models after pioneering studies in this field \cite{wells1934air,riley1978airborne,sze2010review}. 
    

     In this model, we consider an enclosed room of volume $v$ (m$^3$) with a volumetric air exchange rate through the HVAC system $E_{hvac}$ (m$^3$/s). The total number of people in the room is $n$. Each person is assumed to inspire and expire (“exchange”) air at the rate of $Q$ (m$^3$/s) on average. The probability of a person being \emph{initially} infected is $q_i$ and the virion emission rate for an infected person is $QC_i^a$ (virions/s) if the person is being active (e.g., lecturing loudly) and $QC_i^p$ (virions/s) if the person is being passive (e.g., listening quietly).  A virion is a single infectious virus particle. The probability of a person being active is $p_a$; the probability of a person being passive is therefore $p_p=1-p_a$.  We further assume that the use of face coverings with the filtration efficiency $f$ and $\hat{f}$ for inhalation and exhalation, respectively. Finally, we assume that the occupants remain in the enclosed room for the duration of $T$. Using these parameters, we can estimate the average airborne virus concentration $C$ using a room-scale mass balance as:
    \begin{align}
    \label{equ:C0}
        C(n, q_i, p_a) = (1-\hat{f}) \frac{n}{E_{hvac} v} q_i \left( p_a {QC}_i^a + (1-p_a) {QC}_i^p \right)
    \end{align} 
    and estimate the average virus dose $D$ for an occupant as:
    \begin{align}
    \label{equ:D}
        & \nonumber D(n, q_i, p_a) = C(n, q_i, p_a)(1-f)QT = \\
        & (1-\hat{f})(1-f) \frac{n}{E_{hvac} v} q_i \left( p_a {QC}_i^a + (1-p_a) {QC}_i^p \right) QT
    \end{align}
    Note that the above equations assume steady state conditions.  Further, for simplicity, this formulation neglects virus removal due to settling and the decay in the number of viable (or infectious) virus particles over time. In other words, the primary sink of virions is assumed to be air exchange through the HVAC system. This is a reasonable simplification given that (1) settling timescales for aerosols are typically an order of magnitude higher compared to the time scales associated with air turnover \cite{oliveira2021aerosol}, and (2) there is significant variability in estimates for how long SARS-CoV-2 remains viable in aerosols or droplet nuclei \cite{van2020aerosol}.
    
    The virus emission rates $QC_i^a$ and $QC_i^p$ can be estimated based on known virus concentrations and aerosol volumes for typical active and passive activities \cite{asadi2019aerosol,stadnytskyi2020airborne,buonanno2020estimation}. For instance, Stadnytskyi \textit{et al.}~\cite{stadnytskyi2020airborne} estimate that 25s of active or loud speaking leads to the emission of between 60nL and 320nL of oral fluid. The viral load in the sputum is estimated to be $c_v \approx 7 \times 10^6$ virions/cm$^{3}$, though this may be as high as O$(10^9)$ virions/cm$^{3}$ \citep{wolfel2020virological}. Based on these estimates, the virus output for an active infected person is expected to be $Q C_i^a \approx 17-90$ virions/s.  Further, the data presented in Buonanno \textit{et al.}~\cite{buonanno2020estimation} suggest that virus emissions are roughly 40 times higher while speaking when compared to resting conditions. Assuming $QC_i^a/QC_i^p \approx 40$, virus emissions from passive persons are expected to be $QC_i^p \approx 0.4-2.3$ virions/s.
    
    Given the average virus dosage, we can calculate the infection probability for one individual in the room after the exposure to other potentially initially infected occupants as:
    \begin{equation}
    \label{equ:P_infec}
        P_{i} = 1 - e^{-\frac{D(n, q_i, p_a)}{D_0}}
    \end{equation}
    where $D_0$ is the dose that leads to transmission in approximately $63\%$ of cases \cite{riley1978airborne,buonanno2020estimation}. Note that the exponential mapping used to translate virus dose into a transmission probability implicitly accounts for the variation in physiological responses to the same exposure as well as the room-scale variation in exposure that the well-mixed model neglects (i.e., arising from concentration hotspots). To our knowledge, the infectious dose for SARS-CoV-2 remains uncertain, but previous estimates for SARS-CoV-1 and Influenza A suggest that $300-800$ virions are needed to cause infection in $50\%$ of the population~\cite{watanabe2010development,schroder2020covid}. If the infectious dose is $D_0 = 1000$ virions, the respiratory emission estimates provided above suggest that active infectious persons with $QC_i^a \approx 17-90$ virions/s can emit approximately 60 to 320 infectious doses per hour while passive persons can emit approximately 1.8 to 7 infectious doses per hour. These ranges are consistent with the estimates provided by Buonanno \emph{et al.}~\cite{buonanno2020estimation}, who suggest that infectious persons undergoing light activity and talking can generate over 100 quanta per hour, where a quantum is defined as the dose required to cause infection in $63\%$ of susceptible persons.  We recognize that there is significant variability in our estimates for both virus emissions and infectious dose.  As a result, any predictions for \textit{absolute} infection risk must be treated with caution.  Nevertheless, predictions generated using the physics-based model presented in this section should still provide useful estimates for \textit{relative} risk under different scenarios.
    
\vspace{-1mm}    
\section{Classroom Risk Model}
\label{sec:classroom_risk_model}
    In this section, we take the model presented in section \ref{sec:general_risk_model} and adapt it to consider classroom interactions. A classroom is assumed to be occupied by instructors (teachers) and students. Instructors are more likely to be active (i.e., lecturing) during a class while students are more likely to be passive (i.e., listening). Therefore, to better model classrooms interactions, we assume different activity levels for instructors (teachers) and students and we also consider the effect of differing initial infection probabilities for instructors (teachers) and students. Specifically, we assume instructors (teachers) and students have activity rates of $p_a^t$ and $p_a^s$, respectively. Similarly, we assume instructors and students have initial infection probabilities $q_i^t$ and $q_i^s$, respectively. We further assume that we have $N^s$ students in a classroom and 1 instructor. $N^s$ is the number of students attending the class in person and is given by:
    \begin{equation}
        N^s = \alpha n, 
    \end{equation}
    where $n$ is the total number of students enrolled in the class and $\alpha$ is the occupancy ratio of the students who attend the class in person. The average viral dose from $m$ infected students in a classroom can be calculated as:
    \begin{equation}
        D_m^s = D(m, 1, p_a^s) 
    \end{equation}
    and the average viral dose from one infected instructor can be calculated as:
    \begin{equation}
        D^t = D(1, 1, p_a^t).
    \end{equation}
    The infection probability for a student in a given classroom after one session, for the case that the instructor and $m$ students are initially infected, is given by:
    \begin{equation}
        P_{i}(1, m) = 1 - e^{-\frac{D^t+D_m^s}{D_0}}.
    \end{equation}
    The infection probability for a student in the case that the instructor is not initially infected but $m$ students are initially infected is given by:
    \begin{equation}
        P_{i}(0, m) = 1 - e^{-\frac{D_m^s}{D_0}}.
    \end{equation}
    To find the total infection probability for a student in a class session, we have to first compute the probability that $m$ students out of $N^s$ will be infectious given the initial infection probability of 1 student as $q_i^s$. For this purpose, we use the following binomial probability:
    \begin{equation}
        p_i^s(m, N^s) = {N^s \choose m} (q_i^s)^m (1 - q_i^s)^{N^s-m}.
    \end{equation}
    Finally, to calculate the total infection probability for any one student after one class meeting given the initial infection probability for the instructor, $q_i^t$, and the students, $p_i^s(m, N^s)$, we have:
    \begin{equation}
    \label{eq:p_infec_class}
        P_{i, class}^{s} = \sum \limits _{m=0}^N p_i^s(m, N^s) [q_i^t P_{i}(1, m) + (1-q_i^t) P_{i}(0, m) ].
    \end{equation}
    Using the above formulation, we can now formally define three important metrics for assessing the impact of classroom interactions over one week: (i) $P_{i, week}^{s}$, the individual infection probability for students\footnote{For simplicity and ease of exposition, we focus only on students because we assume a high student-faculty ratio at a university.}, (ii) $\hat{P}_{i,week}^s$, the average infection probability, and (iii) $R_0^{eff}$, the effective reproduction number. We define these metrics below.
    
    The reason that we present these metrics with reference to the time period of one week is in part a modeling choice -- our analysis and simulations could easily be carried out for any other time period. However, one week is a natural time-scale to focus on for two reasons.  First, class schedules repeat weekly.  Second, COVID-19 symptoms appear at the latest after two weeks, and on average symptomatic patients show the symptoms after one week~\cite{backer2020incubation}. We assume individuals would not attend classes once they are symptomatic.

    If a particular student $j$ attends $k$ classes with infection probabilities of $p_1, p_2, \cdots, p_k$, the \textbf{individual infection probability} for this particular student after attending one week of classes will be:
    \begin{equation}
    \label{eq:infec_week}
        P_{i, week}^{s}(j) = 1 - (1-p_1)^{n_1} (1-p_2)^{n_2} \cdots (1-p_k)^{n_k},
    \end{equation}
    where $n_i$ is the number of sessions for the class $i$.
    We can then define the \textbf{average infection probability} after one week of classes as:
    \begin{equation}
    \label{eq:infec_average}
        \hat{P}_{i,week}^s =   \frac{1}{N^s}\sum\limits_{j=1}^{N^s} P_{i,week}^{s}(j).
    \end{equation}
    A well-known parameter for infection spread in epidemics is $R_0$, referred to as the {\it expected reproductive number}, which indicates the average number of individuals infected by one initially infected individual in a population. For classroom interactions over a week, we can define a similar ratio of new cases to initial cases by taking the ratio of infection probabilities before and after the week. We thus define the \textbf{effective $R_0$} from one week of operating classes as:
    \begin{equation}
    \label{eq:infec_R0}
        R_0^{eff} = \frac{\hat{P}_{i,week}^s}{q_i^s}.
    \end{equation}

\section{Dataset}
\label{sec:dataset}
    We obtained registration information of all students for a large US university for Fall 2019 and Fall 2020. Both datasets include information for each student registered for classes. For Fall 2019, we consider only classes that were held in person (as most of them were).
    However, the classes in Fall 2020 were either online or hybrid. In hybrid mode, a fraction of students are assumed to attend the class in person and the rest are assumed to watch the class online. As shown in table \ref{table:dataset}, for Fall 2019 there were 5986 courses with 34042 students on campus. For Fall 2020, there were 523 hybrid courses and 6376 students registered for those classes. The remaining classes were entirely online and are therefore not considered in this study.
    
    \begin{table}[ht]
    \centering
        \caption{Datasets Information}
        \begin{tabular}{|c | c | c |} 
         \hline
         Semester & \#In-Person Courses & \#Students on Campus \\
         \hline
         Fall 2019 & 5986 & 34042 \\ 
         \hline
         Fall 2020 & 523 & 6376 \\
         \hline
        \end{tabular}
        \label{table:dataset}
    \end{table}

    We also obtained a dataset containing information about buildings, classroom sizes, ventilation rates, and maximum occupancy (or capacity). This dataset was used to estimate the physical parameters (classroom volume $v$, air change rate $E_{hvac}$, etc.) appearing in equations \eqref{equ:C0}-\eqref{equ:D}.

\section{Simulation Methodology}
\label{sec:simulation_methodology}

    \begin{table}[t]
        \centering
        \caption{Assumed parameter values, unless specified.}
        \label{table:parameters}
        \begin{tabular}{|l|c|}
        \hline
        Occupancy ratio, $\alpha$ & 0.2 \\
        Respiration rate, $Q$ & $10^{-4}$ m$^3$s$^{-1}$ \\
        Active emission rate, $QC_i^a$ & 40 virions s$^{-1}$ \\
        Passive emission rate, $QC_i^p$ & 1 virion s$^{-1}$ \\ 
        Mask filtration efficiency, $f=\hat{f}$ & 0.5 \\
        Infectious Dose, $D_0$ & 1000 virions \\
        Active fraction instructors, $p_a^t$ & 0.9 \\
        Active fraction students, $p_a^s$ & 0.05 \\
        Initial infection probability instructor, $q_i^t$ & 0.01 \\
        Initial infection probability student, $q_i^s$ & 0.01 \\
        \hline
        \end{tabular}
    \end{table}

    We performed simulations to understand the impact of holding one week of classes. We used the three metrics introduced in section~\ref{sec:classroom_risk_model} to evaluate the impact of different policies on virus spread predicted in our simulations.
    The first metric is the infection probability of individual students after attending one week of classes, shown in equation \eqref{eq:infec_week}. This metric is shown via histograms in figure~\ref{fig:hist} for different policies for the Fall 2019 and Fall 2020 datasets.
    The second metric is the average infection probability of the students after attending one week of classes, shown in equation \eqref{eq:infec_average}. 
    The third metric is the effective reproduction number $R_0^{eff}$ for the students given in equation \eqref{eq:infec_R0} for one week of classes. This metric was calculated for differing scenarios and is presented in Table~\ref{table:hist_info}.
    In the simulations, we explicitly study the impact of the following parameters on classroom transmission.
    \begin{itemize}
        \item $\alpha$: This parameter represents the occupancy ratio of the number of students who attend the class in person. We consider the default value for the occupancy ratio to be $\alpha=0.2$ for hybrid classes in the Fall 2020 dataset, which assumed that 20\% of registered students attend the classes in person. For in person classes in the Fall 2019 dataset, we assume full occupancy, $\alpha=1$.
        \item $f, \hat{f}$: These parameters represent how effective masks are in decreasing transmission probability. The default value for the mask efficiency is considered to be 0.5 based on experimental measurements made for a variety of common mask materials \cite{konda2020aerosol}. 
    \end{itemize}
    For reference, default values for each parameter are shown in table \ref{table:parameters}.

\section{Results and Discussion}
\label{sec:results_discussion}

    We now study the impact of different policies on virus spread for the Fall 2019 and Fall 2020 datasets in the context of the three different metrics discussed in section \ref{sec:simulation_methodology}.

    \begin{table*}[t]
    \caption{Infection Probability Distribution Information}
    \centering
    \begin{tabular}{|l|c|c|c|c|c|c|}
    \hline
    \multirow{2}{*}{\textbf{}}                                              & \multicolumn{3}{c|}{Fall 2019}                                                                                                                                                          & \multicolumn{3}{c|}{Fall 2020}                                                                                                                                                          \\ \cline{2-7} 
                                                                            & \begin{tabular}[c]{@{}c@{}}$\alpha$ = 1\\ $f=\hat{f}$ = 0\end{tabular} & \begin{tabular}[c]{@{}c@{}}$\alpha$ = 1\\ $f=\hat{f}$ = 0.5\end{tabular} & \begin{tabular}[c]{@{}c@{}}$\alpha$ = 0.2\\ $f=\hat{f}$ = 0.5\end{tabular} & \begin{tabular}[c]{@{}c@{}}$\alpha$ = 1\\ $f=\hat{f}$ = 0\end{tabular} & \begin{tabular}[c]{@{}c@{}}$\alpha$ = 1\\ $f=\hat{f}$ = 0.5\end{tabular} & \begin{tabular}[c]{@{}c@{}}$\alpha$ = 0.2\\ $f=\hat{f}$ = 0.5\end{tabular} \\ \hline
    \begin{tabular}[c]{@{}l@{}}Average Infection Probability\end{tabular} & 0.05407                                                   & 0.01480                                                     & 0.00634                                                       & 0.01608                                                   & 0.00438                                                     & 0.00204                                                       \\ \hline
    \begin{tabular}[c]{@{}l@{}}Average New Infected Students\end{tabular} & 1864                                                      & 504                                                         & 216                                                           & 103                                                       & 28                                                          & 13                                                            \\ \hline
    $R_0^{eff}$                                                                   & 5.407                                                     & 1.480                                                        & 0.634                                                         & 1.608                                                     & 0.438                                                       & 0.204                                                         \\ \hline
    \end{tabular}
    \label{table:hist_info}
    \end{table*}

    \begin{figure*}
        \centering 
    \begin{subfigure}[b]{\columnwidth}
      \includegraphics[width=\linewidth]{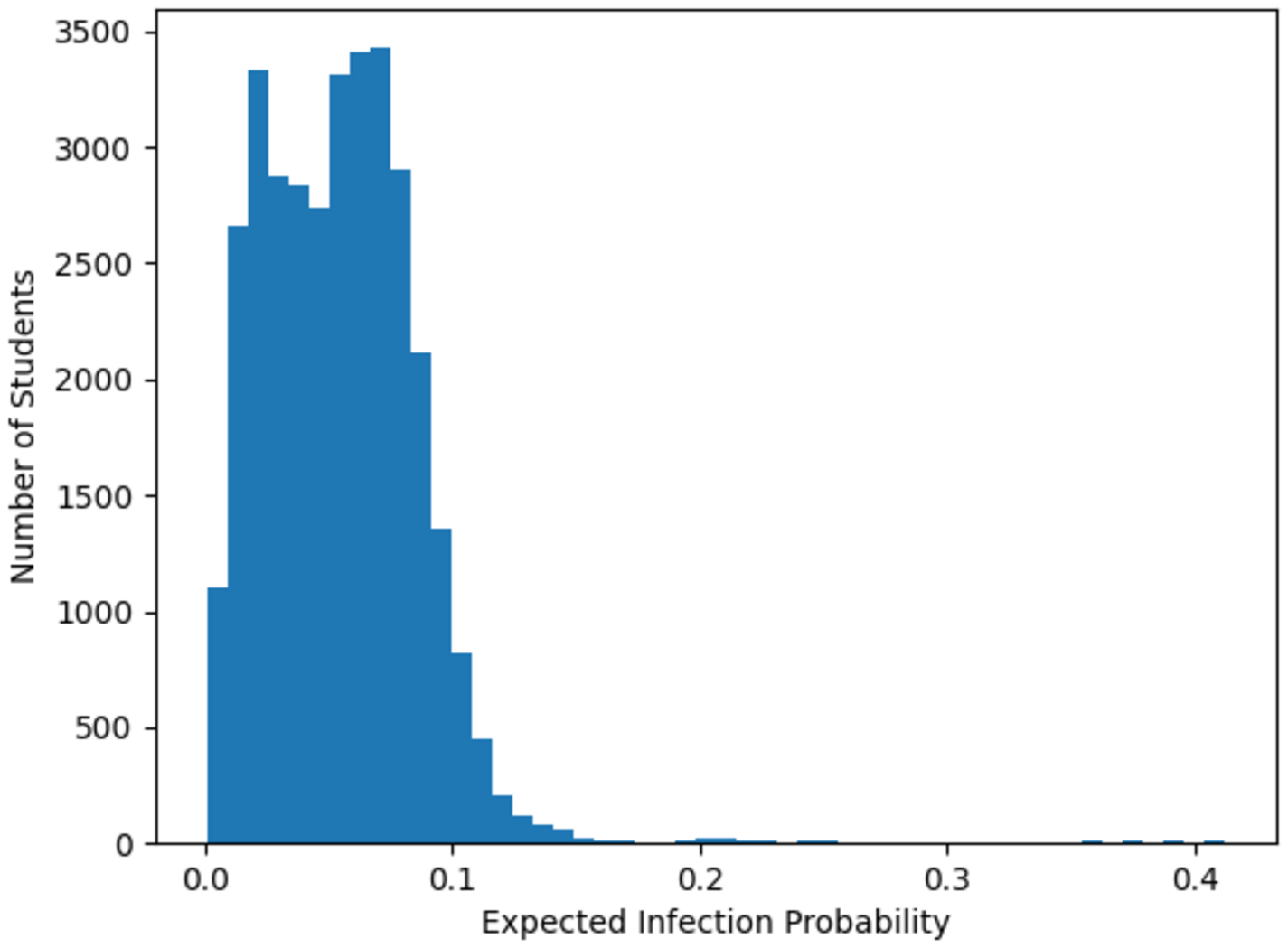}
      \caption{Fall 2019, $\alpha$= 1, $f=\hat{f}$= 0, $\hat{P}_{i,week}^s$= 0.05407}
      \label{fig:hist_2019_ocp_1_msk_0}
    \end{subfigure}\hfill
    \begin{subfigure}[b]{\columnwidth}
      \includegraphics[width=\linewidth]{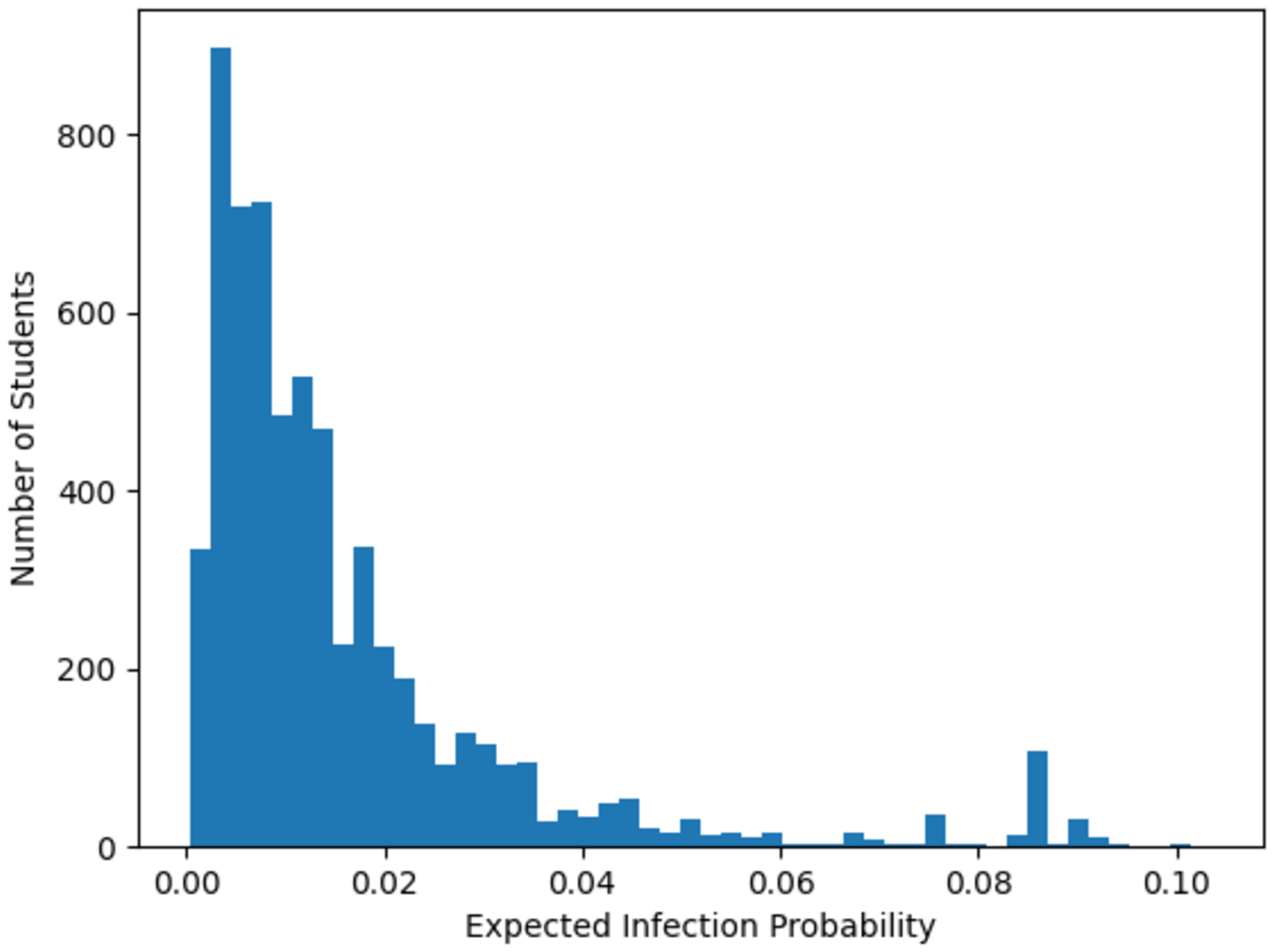}
      \caption{Fall 2020, $\alpha$= 1, $f=\hat{f}$= 0, $\hat{P}_{i,week}^s$= 0.01608}
      \label{fig:hist_2020_ocp_1_msk_0}
    \end{subfigure}

    \medskip
    \begin{subfigure}[b]{\columnwidth}
      \includegraphics[width=\linewidth]{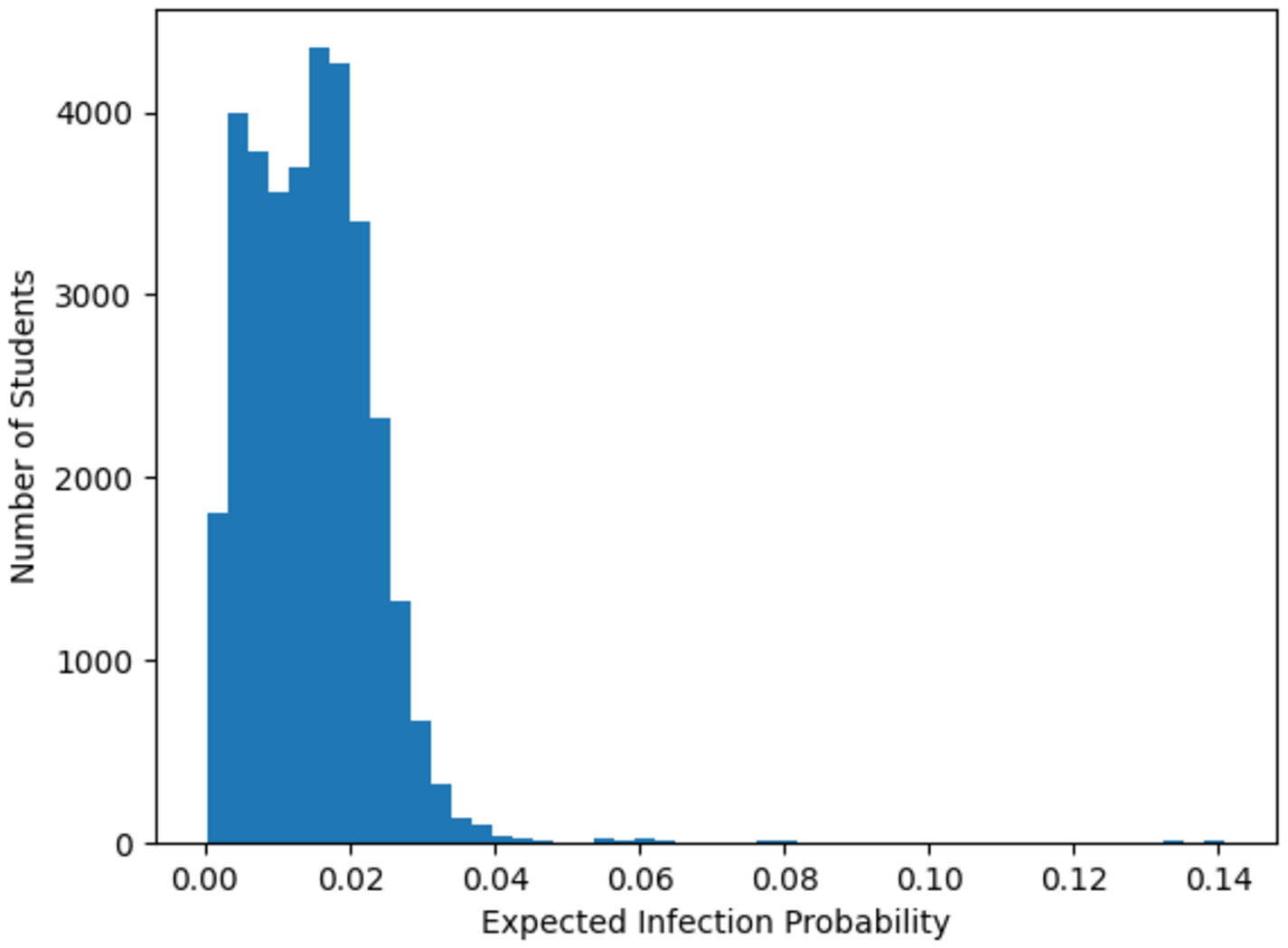}
      \caption{Fall 2019, $\alpha$= 1, $f=\hat{f}$= 0.5, $\hat{P}_{i,week}^s$= 0.01480}
      \label{fig:hist_2019_ocp_1_msk_0.5}
    \end{subfigure}\hfill 
    \begin{subfigure}[b]{\columnwidth}
      \includegraphics[width=\linewidth]{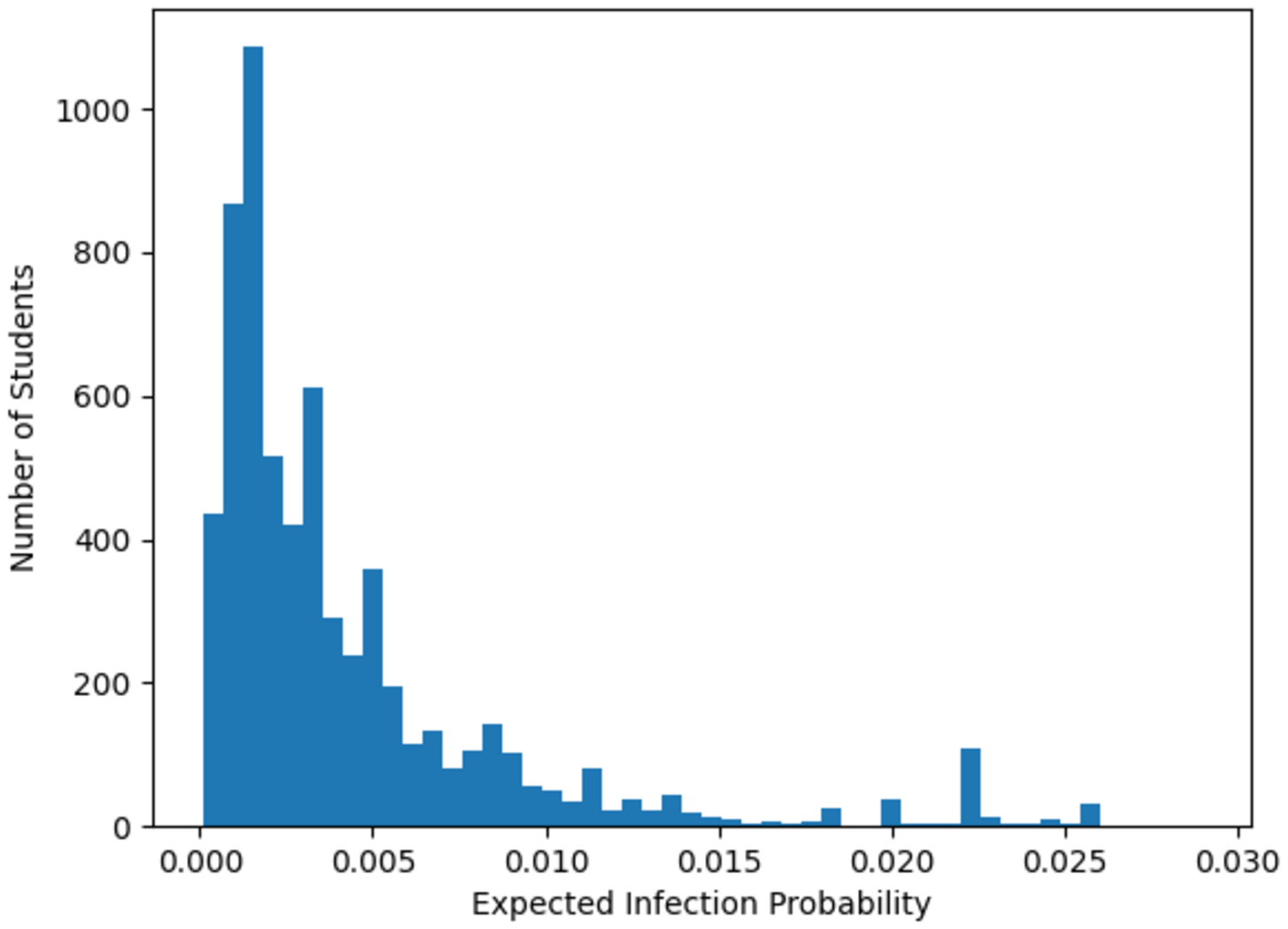}
      \caption{Fall 2020, $\alpha$= 1, $f=\hat{f}$= 0.5, $\hat{P}_{i,week}^s$= 0.00438}
      \label{fig:hist_2020_ocp_1_msk_0.5}
    \end{subfigure}

    \medskip
    \begin{subfigure}[b]{\columnwidth}
      \includegraphics[width=\linewidth]{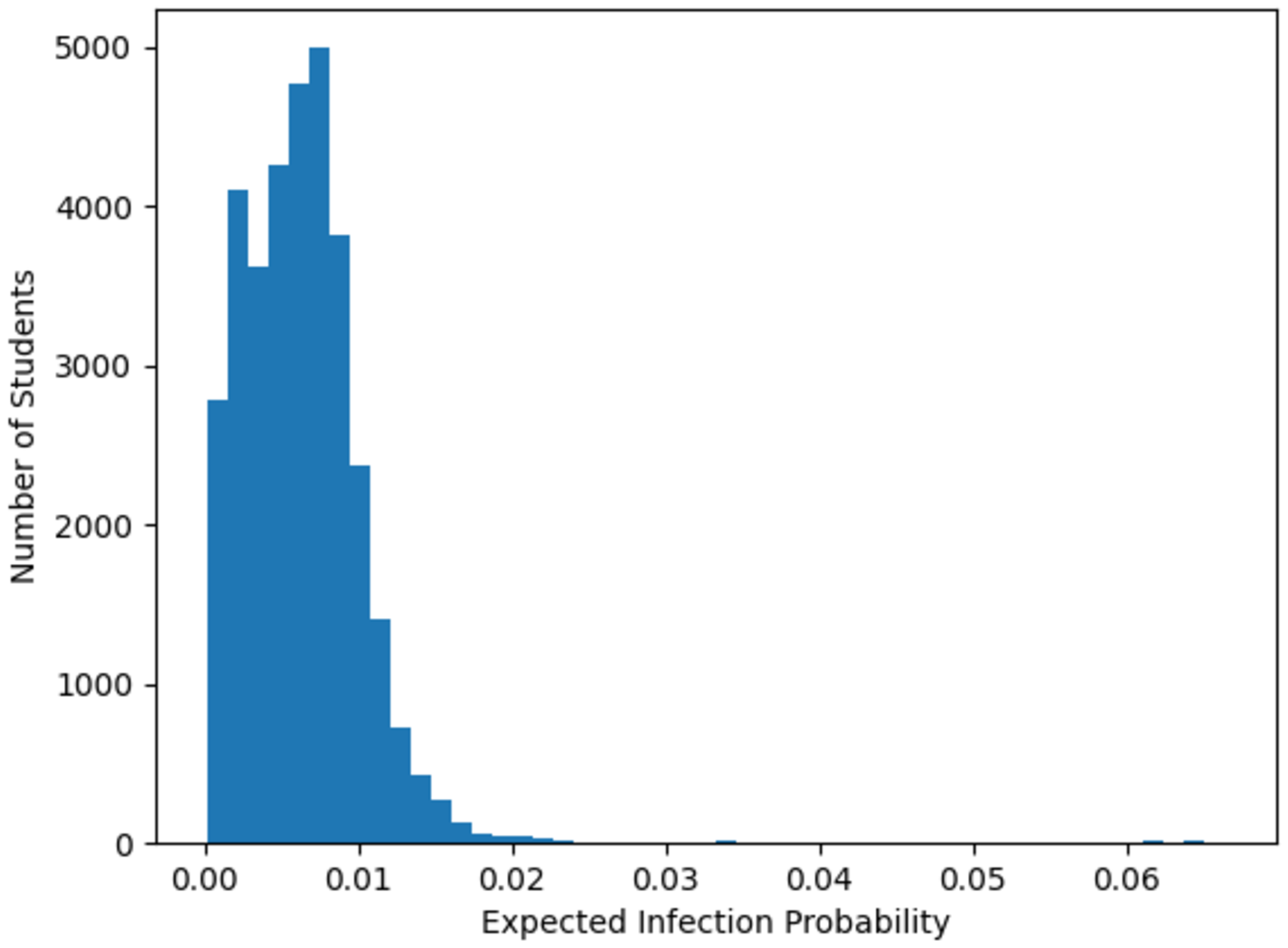}
      \caption{Fall 2019, $\alpha$= 0.2, $f=\hat{f}$= 0.5, $\hat{P}_{i,week}^s$= 0.00634}
      \label{fig:hist_2019_ocp_0.2_msk_0.5}
    \end{subfigure}\hfill 
    \begin{subfigure}[b]{\columnwidth}
      \includegraphics[width=\linewidth]{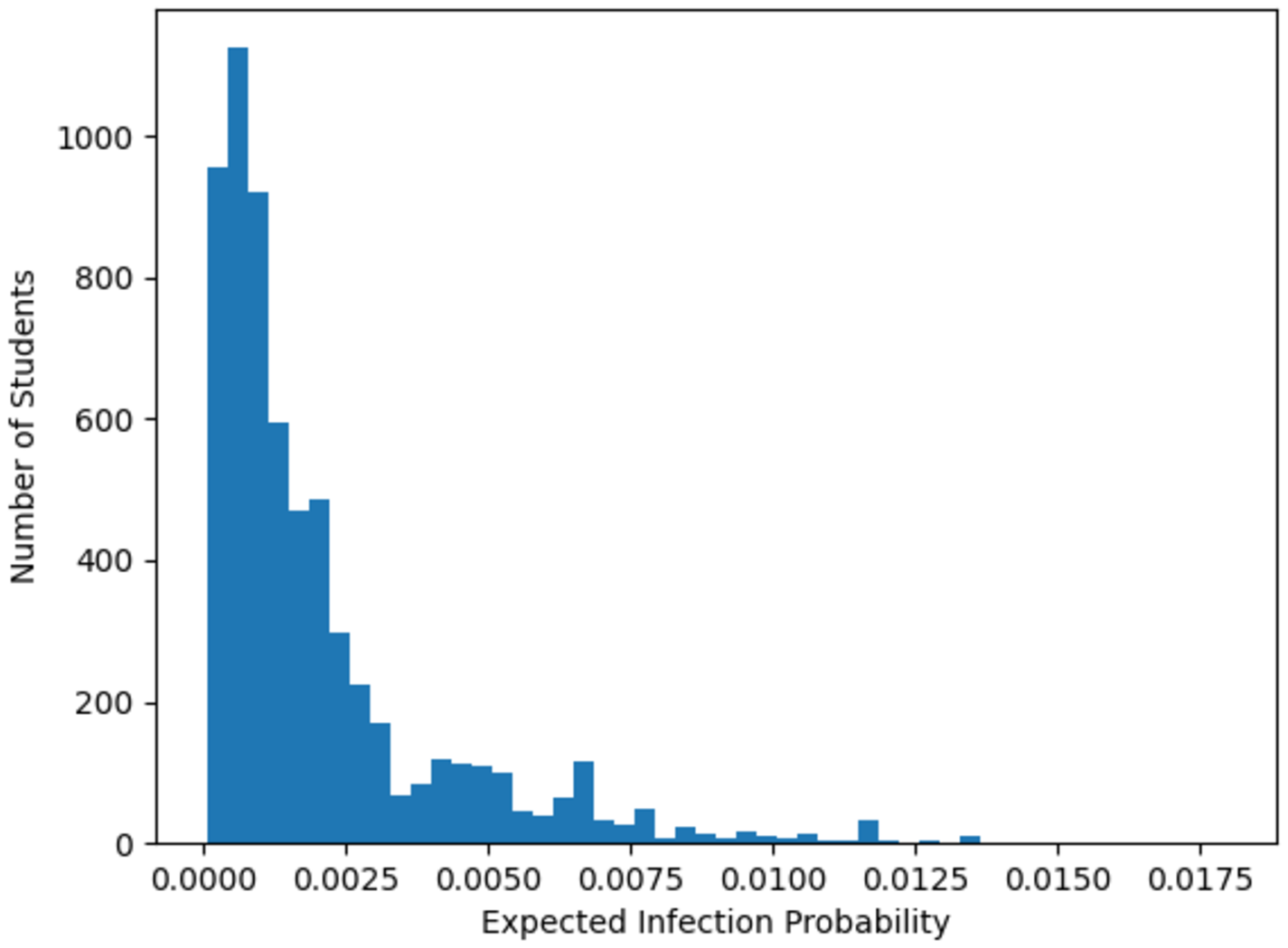}
      \caption{Fall 2020, $\alpha$= 0.2, $f=\hat{f}$= 0.5, $\hat{P}_{i,week}^s$= 0.00204}
      \label{fig:hist_2020_ocp_0.2_msk_0.5}
    \end{subfigure}
    \caption{Infection probability distributions for the students after attending one week of classes, assuming initial infection probabilities of 0.01 for both students and instructors, i.e., $q_i^s$ = $q_i^t$ = 0.01. The average infection probability ($\hat{P}_{i,week}^s$) after one week of classes is shown in the individual captions.}
    \label{fig:hist}        
    \end{figure*}

    Figure~\ref{fig:hist} shows infection probability distributions for the students after one week of classes under three different occupancy and masking scenarios for Fall 2019 and Fall 2020.  For the business-as-usual case with the Fall 2019 dataset (i.e., in-person classes, full occupancy, no mask usage), figure~\ref{fig:hist_2019_ocp_1_msk_0} shows that a significant proportion of the students have high individual infection probabilities, $P_{i, week}^{s} \ge 0.05$.  Indeed, the average infection probability after one week of classes is $\hat{P}_{i,week}^s = 0.05407$, which is more than five times higher than the initial infection probabilities for the students and instructors.  In other words, the simulations predict that holding classes in person without occupancy reductions or masking requirements would have led to a substantial increase in infection rates for the students.  
    
    Figure~\ref{fig:hist_2020_ocp_1_msk_0} shows that the transition to online instruction with limited hybrid classes leads to a substantial reduction in infection probabilities compared to the business-as-usual scenario.  Specifically, the average infection probability for the students reduces by more than a factor of 3 to $\hat{P}_{i,week}^s = 0.01508$ even without occupancy reductions or mask usage.  Note that this average infection probability is still higher than the initial infection probabilities at the start of the week, which is indicative of a reproductive number greater than 1.  However, as shown in figures~\ref{fig:hist_2020_ocp_1_msk_0.5} and \ref{fig:hist_2020_ocp_0.2_msk_0.5}, the use of $50\%$ effective masks and a reduction to $20\%$ occupancy leads to significant further reductions in average infection probabilities to levels where the effective reproductive number is less than 1.  

    Table \ref{table:hist_info} provides a summary of average infection probabilities and effective reproductive numbers for the different scenarios we considered.  The simulations predict a lower average infection probability for Fall 2020 compared to Fall 2019 across all masking and occupancy conditions. They indicate that universal mask usage results in an approximately $3.6\times$ reduction in new infections through classroom interactions. In addition, reducing class occupancy to 20\%,  by having hybrid classes, results in an approximately $2.15-2.3\times$ further reduction in new infections.
    The transition to having 90\% of the courses online between Fall 2019 to Fall 2020 alone, even without using masks or reducing the occupancy, results in a 18$\times$ reduction of cases.

\section{Conclusion}
\label{sec:conclusion}
    In this work, we have studied the impact of various policies on COVID-19 transmission via classroom interactions at  universities. Our specific aim has been to quantify the effect of different policies on transmission rates, and thereby enable institutions of higher education to prepare for future epidemics. We utilized an established model for airborne transmission in enclosed classrooms and registration data from a large university to perform simulations for different scenarios. We also introduced a quantitative metric $R_0^{eff}$ that represents the ratio of the infection probability after holding one week of classes to the initial infection probability. We consider this metric to be analogous to the well-known reproductive number, $R_0$. 
    We have analyzed the effect of classroom occupancy, mask usage, and initial infection probabilities in the student and instructor populations on transmission. The simulations also accounted for repeated interactions in classes between students throughout the week. In the future, it would be of interest to also investigate the effect of student interactions outside of classes, although, it is challenging to find relevant data.

\section{Acknowledgments}
We are grateful to the collaborators who provided the university data used in this work.

\bibliographystyle{IEEEtran}
\bibliography{main.bib}

\begin{thebibliography}{10}
\providecommand{\url}[1]{#1}
\csname url@samestyle\endcsname
\providecommand{\newblock}{\relax}
\providecommand{\bibinfo}[2]{#2}
\providecommand{\BIBentrySTDinterwordspacing}{\spaceskip=0pt\relax}
\providecommand{\BIBentryALTinterwordstretchfactor}{4}
\providecommand{\BIBentryALTinterwordspacing}{\spaceskip=\fontdimen2\font plus
\BIBentryALTinterwordstretchfactor\fontdimen3\font minus
  \fontdimen4\font\relax}
\providecommand{\BIBforeignlanguage}[2]{{%
\expandafter\ifx\csname l@#1\endcsname\relax
\typeout{** WARNING: IEEEtran.bst: No hyphenation pattern has been}%
\typeout{** loaded for the language `#1'. Using the pattern for}%
\typeout{** the default language instead.}%
\else
\language=\csname l@#1\endcsname
\fi
#2}}
\providecommand{\BIBdecl}{\relax}
\BIBdecl

\bibitem{nytimes2020track}
``{Tracking Covid at U.S. Colleges and Universities},''
  https://www.nytimes.com/interactive/2020/us/covid-college-cases-tracker.html,
  accessed: 2020-10-12.

\bibitem{education2020closure}
``{Map: Coronavirus and School Closures in 2019-2020},''
  https://www.edweek.org/ew/section/multimedia/map-coronavirus-and-school-closures.html,
  accessed: 2020-10-12.

\bibitem{duong2020ivory}
V.~Duong, P.~Pham, T.~Yang, Y.~Wang, and J.~Luo, ``{The ivory tower lost: How
  college students respond differently than the general public to the COVID-19
  pandemic},'' \emph{arXiv preprint arXiv:2004.09968}, 2020.

\bibitem{wiki2020impact}
``{Impact of the {COVID-19} pandemic on education},''
  https://en.wikipedia.org/wiki/Impact\_of\_the\_{COVID-19}\_pandemic\_on\_education,
  accessed: 2020-10-12.

\bibitem{nazaroff2004indoor}
W.~W. Nazaroff, ``{Indoor particle dynamics},'' \emph{Indoor air}, vol.~14, no.
  Supplement 7, pp. 175--183, 2004.

\bibitem{nazaroff2016indoor}
------, ``{Indoor bioaerosol dynamics},'' \emph{Indoor Air}, vol.~26, no.~1,
  pp. 61--78, 2016.

\bibitem{wells1934air}
W.~Wells, ``{ON air-borne infection: Study II. Droplets and droplet nuclei.}''
  \emph{American journal of Epidemiology}, vol.~20, no.~3, pp. 611--618, 1934.

\bibitem{riley1978airborne}
E.~Riley, G.~Murphy, and R.~Riley, ``{Airborne spread of measles in a suburban
  elementary school},'' \emph{American journal of epidemiology}, vol. 107,
  no.~5, pp. 421--432, 1978.

\bibitem{sze2010review}
G.~N. Sze~To and C.~Y.~H. Chao, ``{Review and comparison between the
  Wells--Riley and dose-response approaches to risk assessment of infectious
  respiratory diseases},'' \emph{Indoor air}, vol.~20, no.~1, pp. 2--16, 2010.

\bibitem{oliveira2021aerosol}
P.~M. de~Oliveira, L.~C.~C. Mesquita, S.~Gkantonas, A.~Giusti, and
  E.~Mastorakos, ``{Evolution of spray and aerosol from respiratory releases:
  theoretical estimates for insight on viral transmission},'' \emph{Proceedings
  of the Royal Society A: Mathematical, Physical and Engineering Sciences},
  vol. 477, no. 2245, p. 20200584, 2021.

\bibitem{van2020aerosol}
N.~Van~Doremalen, T.~Bushmaker, D.~H. Morris, M.~G. Holbrook, A.~Gamble, B.~N.
  Williamson, A.~Tamin, J.~L. Harcourt, N.~J. Thornburg, S.~I. Gerber
  \emph{et~al.}, ``{Aerosol and surface stability of SARS-CoV-2 as compared
  with SARS-CoV-1},'' \emph{New England Journal of Medicine}, vol. 382, no.~16,
  pp. 1564--1567, 2020.

\bibitem{asadi2019aerosol}
S.~Asadi, A.~S. Wexler, C.~D. Cappa, S.~Barreda, N.~M. Bouvier, and W.~D.
  Ristenpart, ``{Aerosol emission and superemission during human speech
  increase with voice loudness},'' \emph{Scientific reports}, vol.~9, no.~1,
  pp. 1--10, 2019.

\bibitem{stadnytskyi2020airborne}
V.~Stadnytskyi, C.~E. Bax, A.~Bax, and P.~Anfinrud, ``{The airborne lifetime of
  small speech droplets and their potential importance in SARS-CoV-2
  transmission},'' \emph{Proceedings of the National Academy of Sciences}, vol.
  117, no.~22, pp. 11\,875--11\,877, 2020.

\bibitem{buonanno2020estimation}
G.~Buonanno, L.~Stabile, and L.~Morawska, ``{Estimation of airborne viral
  emission: quanta emission rate of SARS-CoV-2 for infection risk
  assessment},'' \emph{Environment International}, p. 105794, 2020.

\bibitem{wolfel2020virological}
R.~W{\"o}lfel, V.~M. Corman, W.~Guggemos, M.~Seilmaier, S.~Zange, M.~A.
  M{\"u}ller, D.~Niemeyer, T.~C. Jones, P.~Vollmar, C.~Rothe \emph{et~al.},
  ``{Virological assessment of hospitalized patients with COVID-2019},''
  \emph{Nature}, vol. 581, no. 7809, pp. 465--469, 2020.

\bibitem{watanabe2010development}
T.~Watanabe, T.~A. Bartrand, M.~H. Weir, T.~Omura, and C.~N. Haas,
  ``{Development of a dose-response model for SARS coronavirus},'' \emph{Risk
  Analysis: An International Journal}, vol.~30, no.~7, pp. 1129--1138, 2010.

\bibitem{schroder2020covid}
I.~Schroder, ``{COVID-19: a risk assessment perspective},'' \emph{ACS Chemical
  Health {\&} Safety}, vol.~27, no.~3, pp. 160--169, 2020.

\bibitem{backer2020incubation}
J.~A. Backer, D.~Klinkenberg, and J.~Wallinga, ``{Incubation period of 2019
  novel coronavirus (2019-nCoV) infections among travellers from {Wuhan},
  {China}, 20-28 January 2020},'' \emph{Eurosurveillance}, vol.~25, 2020.

\bibitem{konda2020aerosol}
A.~Konda, A.~Prakash, G.~A. Moss, M.~Schmoldt, G.~D. Grant, and S.~Guha,
  ``{Aerosol filtration efficiency of common fabrics used in respiratory cloth
  masks},'' \emph{ACS nano}, vol.~14, no.~5, pp. 6339--6347, 2020.

\end{thebibliography}

\end{document}